\documentstyle[amsmath,amssymb,prb,aps,twocolumn,epsfig]{revtex}

\begin{document}
\draft
\title{Ginzburg-Landau theory of superconductors with short 
coherence length}
\author{S. Stintzing and W. Zwerger}
\address{Sektion Physik, Ludwig-Maximilians-Universit\"at M\"unchen,
Theresienstra{\ss}e 37, D-80333 M\"unchen, Germany}
\date{\today}
\maketitle

\begin{abstract}
We consider Fermions in two dimensions with an attractive interaction 
in the singlet d-wave channel of arbitrary strength. By means of a 
Hubbard-Stratonovich
transformation a statistical Ginzburg-Landau theory  is derived, which
describes the smooth crossover from a weak-coupling BCS superconductor
to a condensate of composite Bosons. 
Adjusting the interaction strength
to the observed slope of $H_{c_2}\/$ at  
$T_c\/$ in the optimally doped high-$T_c\/$
compounds YBCO and BSCCO, we determine the associated 
values of the Ginzburg-Landau
correlation length $\xi\/$ and the London penetration depth 
$\lambda\/$. The resulting dimensionless ratio $k_F\xi(0)\approx 5-8\/$
and the Ginzburg-Landau parameter $\kappa=\lambda /\xi \approx 90-100\/$
agree well with the experimentally observed values. These parameters indicate 
that the optimally doped materials are still on the weak coupling side of the 
crossover to a Bose regime. 
\end{abstract}
\pacs{74.20.De}

\section{Introduction}
The problem of the crossover from a BCS-like super\-conducting state 
to a Bose condensate of 
local pairs \cite{Legg80,NozSch84,Rand94} has gained
new interest in the context of high-$T_c\/$ super\-conductors. While
there is still no \mbox{quantitative}
microscopic theory of how superconductivity
arises from doping the antiferromagnetic and insulating parent com\-pounds
\cite{Scalap95}, it is clear that the superconducting state can be 
described in terms of a generalized pairing picture.
The many body ground state is thus a coherent superposition of two particle
states built from spin singlets in a relative d-wave configuration
\cite{AnGoLe96}. The short coherence length $\xi (0)\approx 10-20\;
\mbox{\AA}\/$ parallel to the basic ${\rm CuO_2}$-planes 
which is of the same
order than the average interparticle spacing $k_F^{\,-1}\/$, however
indicates that neither the BCS picture of highly overlapping pairs nor
a description in terms of composite Bosons is applicable here. It is
therefore of considerable interest to develop a theory, which is able to 
cover the whole regime between weak and strong coupling in a unified manner.
On a phenomenological level such a description is provided by the
Ginzburg-Landau (GL) theory. Indeed, it is a nonvanishing
expectation value of the complex order parameter $\psi\/$ which signals
the breaking of gauge invariance as the basic characteristic of the
superconducting state, irrespective of whether the pair size is much larger 
or smaller than the interparticle spacing. A GL description of the
BCS to Bose crossover was developed for the s-wave case by Drechsler 
and one of the present authors \cite{DreZwe92} in two dimensions and by
S\'{a} de Melo, Randeria and Engelbrecht \cite{SaRaEn93} for the 
three-dimensional case. 
In our present work the theory for the two-dimensional case is reconsidered,
including a discussion of the Nelson-Kosterlitz jump of the order
parameter and the generalization
to the experimentally relevant situation of d-wave superconductivity.
Moreover we also calculate the characteristic lengths $\xi\/$ and $\lambda\/$
and compare our results with measured properties of 
high-$T_c\/$ compounds.

The remarkable success with which the standard BCS model has been applied to 
conventional superconductors relies on the fact that in the weak coupling 
limit the details of the attractive interaction are irrelevant. By an 
appropriate rescaling of the parameters the properties of all weak
coupling superconductors are therefore universal. It is one of our aims here to
investigate to which extent such a simplifying description also exists in
more strongly coupled superconductors. 
Starting from a microscopic model with an instantaneous attractive interaction,
we find that the resulting GL functional takes the standard form for
arbitrary strength of the coupling. By adjusting a single dimensionless
parameter to the measured upper critical field near $T_c\/$,
we obtain consistent values for both the dimensionless
ratio $k_F\xi (0)\approx 5-8\/$ between the coherence length and interparticle
spacing as well as the observed value $\kappa\approx 90-100\/$ of the
GL parameter 
in optimally doped high-$T_c\/$ compounds. Therefore, in spite of the
rather crude nature of the original microscopic model, our GL theory is
quantitatively applicable to strongly coupled superconductors which are
far from the standard weak coupling limit, although not yet in the
crossover regime to Bose-like behaviour.

The plan of the paper is as follows: in section II we introduce our 
microscopic model which has an attractive interaction in the singlet 
d-wave channel. From this we derive, via a Hubbard-Stratonovich
transformation, a statistical GL theory which is valid near $T_c\/$.
The relevant coefficients of the GL functional are calculated for
arbitrary interaction strength. 
In section III we discuss the appropriate microscopic definition
of the order parameter and the evolution from the BCS to the Bose limit
of the well known Kosterlitz-Thouless jump in the superfluid density of
two-dimensional superconductors.
Ignoring the subtleties of the Kosterlitz-Thouless transition,
in section IV we use a Gaussian
approximation to determine the critical temperature and the
associated value of the chemical potential at the transition. Finally in
section V we determine the
characteristic lengths $\xi\/$ and $\lambda\/$ near the transition
for arbitrary strength of the coupling.
Adjusting the coupling to the experimental values of 
the slope of ${H_c}_2\/$ near
$T_c\/$, we then determine
the associated dimensionless ratios $k_F\xi (0)\/$ and $\kappa=\xi/
\lambda\/$. They agree rather well with the observed values in YBCO and
BSCCO. A brief conclusion and a discussion of open problems is given
in section VI.

\section{Microscopic derivation of the GL functional}
As a general model describing Fermions with an instaneous pairwise
interaction $V(\boldsymbol{r})\/$ in a translationally invariant system,
we start from the Hamiltonian ($V\/$ is the volume of the system.)
\begin{multline}
H = \sum\limits_{\boldsymbol{k}}\sum\limits_{\sigma}\epsilon_k^{}\,
c_{\boldsymbol{k}\,\sigma}^{\dagger} c_{\boldsymbol{k}\,\sigma}^{} 
\\ +\,\frac{1}{2V}
\sum\limits_{\boldsymbol{k}\,\boldsymbol{k'}\boldsymbol{q}}
\sum\limits_{\sigma\,\sigma'} V_{\boldsymbol{k}-\boldsymbol{k'}}^{}\;
c_{\boldsymbol{k}+\boldsymbol{q}\,\sigma}^{\dagger} 
c_{-\boldsymbol{k}\,\sigma'}^{\dagger} c_{-\boldsymbol{k'}\,\sigma'}^{} 
c_{\boldsymbol{k'}+\boldsymbol{q}\,\sigma}^{} 
\end{multline}
with an arbitrary single particle energy $\epsilon_k^{}\/$ which we will
later replace by an effective mass approximation 
$\epsilon_k^{} = \hbar^2 k^2/2m\/$.
In the two-dimensional case, which we consider throughout, the Fourier
transform $V_{\boldsymbol{k}-\boldsymbol{k'}}^{}\/$ of the interaction
potential may be expanded in its relative angular momentum contributions
$l\/$ by
\begin{equation} 
V_{\boldsymbol{k}-\boldsymbol{k'}}^{} =
V_0^{}(k,k') + 2 \sum\limits_{l=1}^{\infty} 
\cos(l\,\varphi)\,V_l^{}(k,k')
\end{equation}
with $\varphi\/$ the angle between $\boldsymbol{k}\/$ and 
$\boldsymbol{k'}\/$. In the following we are only interested
in d-wave pairs with symmetry ${\rm d_{x^2-y^2}}\/$.
We therefore omit all contributions $l\neq 2\/$ and also neglect the
dependence on the absolute values $k\/$ and $k'\/$ of the momenta. 
Assuming the interaction is separable, we thus approximate 
\begin{equation} \label{ref3}
V_{\boldsymbol{k}-\boldsymbol{k'}}^{} \rightarrow -g\;
v_{\boldsymbol{k}+\boldsymbol{q}/2}^{}\,
v_{\boldsymbol{k'}+\boldsymbol{q}/2}^{}
\end{equation}
with  
\begin{equation} 
v_{\boldsymbol{k}}^{} = \sqrt{2}
\,\frac{k_y^{\,2}-k_x^{\,2}}{k^2}
\end{equation}
and $g\/$ a negative constant characterizing the strength of the
attractive interaction. Finally the 
restriction to singlet pairing is incorporated trivially by considering only 
interactions between Fermions with opposite spins $\sigma'=-\sigma\/$.
In this manner we obtain a Gorkov-like reduced interaction Hamiltonian
\begin{multline} 
H' = -\frac{g}{2V}
\sum\limits_{\boldsymbol{k}\,\boldsymbol{k'}\,\boldsymbol{q}}
\sum\limits_{\sigma} v_{\boldsymbol{k}+\boldsymbol{q}/2}^{}\,
v_{\boldsymbol{k'}+\boldsymbol{q}/2}^{}\;\\ \cdot\,
c_{\boldsymbol{k}+\boldsymbol{q}\,\sigma}^{\dagger} \,
c_{-\boldsymbol{k}\,-\sigma}^{\dagger}\, c_{-\boldsymbol{k'}\,-\sigma}^{}\, 
c_{\boldsymbol{k'}+\boldsymbol{q}\,\sigma}^{}
\end{multline}
(Note that the shift by $\boldsymbol{q}/2\/$ in Eq. (\ref{ref3}) is 
necesssary to guarantee that the interaction is symmetric with respect to 
$\sigma\leftrightarrow-\sigma\/$.). 
For the derivation of a GL functional below, it is 
convenient to introduce pair operators $b_{\boldsymbol{q}}\/$ via 
\begin{equation} 
b_{\boldsymbol{q}}^{} = \sum\limits_{\boldsymbol{k}} 
v_{\boldsymbol{k}+\boldsymbol{q}/2}^{}\;
c_{-\boldsymbol{k}\;+1}^{}\,
c_{\boldsymbol{k}+\boldsymbol{q}\;-1}^{}\;. 
\end{equation}
The contribution $H'\/$ may then be written in the form
\begin{equation} 
H' = -\frac{g}{V}\sum\limits_{\boldsymbol{q}}
b_{\boldsymbol{q}}^{\dagger}\,b_{\boldsymbol{q}}^{}
\end{equation}
of an attractive interaction between pairs of Fermions with 
opposite spin and total momentum $\boldsymbol{q}\/$. 
In the following we want to derive
a functional integral representation of the grand partition function
\begin{equation} 
Z = {\rm Tr}\;e^{-\beta(H-\mu N)}
\end{equation}
which gives the standard GL theory as its mean field limit. Since we 
are interested in a superconducting state with a nonzero anomalous average
$\langle b_{\boldsymbol{q}}^{}\rangle \neq 0\/$, it is convenient to
formally linearize the interaction term $H'\/$ by a Hubbard-Stratonovich
transformation \cite{Mueh77}. The grand partition function is thus 
expressed in terms of a functional integral
\begin{equation} \label{ref9} \hspace*{-0.1in}
Z = {\displaystyle \int} D^2 z\;\exp\bigg(-\frac{1}{V\hbar g}
\int\limits_{0}^{\beta\hbar} d\tau \sum\limits_{\boldsymbol{q}}
|z(\boldsymbol{q},\tau)|^2 \bigg) \;L[z]  
\end{equation}
over a complex valued c-number field 
$z(\boldsymbol{q},\tau)\/$. Here
\begin{equation} \label{ref10}
L[z] = {\rm Tr}\;\,{\rm T}\exp
\bigg( -\frac{1}{\hbar}\int\limits_0^{\beta\hbar}d\tau\;
H_z(\tau) \bigg)
\end{equation}
is a functional of the auxiliary field $z(\boldsymbol{q},\tau)\/$,
which acts as a space- and 'time'-dependent external potential on a
noninteracting Fermi system with Hamiltonian
\begin{multline} 
H_z (\tau) = \sum\limits_{\boldsymbol{k}}\sum\limits_{\sigma}
(\epsilon_{k}^{}-\mu)\,
c_{\boldsymbol{k}\,\sigma}^{\dagger} c_{\boldsymbol{k}\,\sigma}^{} \\
+\,\frac{1}{V}\sum\limits_{\boldsymbol{q}} \big(z(\boldsymbol{q},\tau)\,
b_{\boldsymbol{q}}^{\dagger} +z^{\star}(\boldsymbol{q},\tau)\,
b_{\boldsymbol{q}}^{}\big) \;.
\end{multline}
The physical interpretation of the c-number field 
$z(\boldsymbol{q},\tau)\/$ is obtained by noting that its expectation value
\begin{equation} \label{expect}
\langle z(\boldsymbol{q},\tau)\rangle = -g 
\langle b_{\boldsymbol{q}}(\tau)\rangle
\end{equation}
is directly proportional to the anomalous average 
$\langle b_{\boldsymbol{q}}(\tau)\rangle\/$. Thus
up to some normalization constant, which will be determined below, the field
$z(\boldsymbol{q},\tau)\/$ is just the spatial Fourier transform of
the complex order parameter $\psi(\boldsymbol{r},\tau)\/$ describing the 
superconducting state. It depends both on position and imaginary time
$\tau\in [0,\beta\hbar]\/$ which is characteristic for a quantum
GL functional. Since the ${\rm d_{x^2-y^2}}\/$-symmetry
in a rotational invariant system is connected with a one dimensional
irreducible representation \cite{SigUed91}, the order parameter is still
a simple complex scalar, similar to the more familiar isotropic s-wave
case. 

Obviously the trace in the time ordered exponential in Eq. (\ref{ref10})
cannot be calculated exactly. However it is straightforward to evaluate
$L[z]\/$ perturbatively in $z\/$. The naive justification for this is that
close to $T_c\/$ the order parameter is small. Strictly speaking however,
the functional integral in Eq. (\ref{ref9}) requires to integrate 
over arbitrary 
realizations of $z(\boldsymbol{q},\tau)\/$. In order to
obtain the standard form of the statistical GL functional, however, the
expansion is truncated at fourth order in the exponent of $L[z]\/$. 
In the language of field theory we
are therefore calculating the bare coupling constants, which serve as the
starting point for treating the behaviour at long wavelenghts. By
a straightforward perturbative calculation \cite{stin96} up to fourth order
in $z\/$, the functional $L[z]\/$ turns out to be of the form
\begin{multline} 
L[z] = Z_0\;\exp\bigg(\frac{1}{V\beta\hbar^2}\sum\limits_{\boldsymbol{q}}
\sum\limits_{\omega_n} \tilde{a}(\boldsymbol{q},\omega_n)\; 
|z(\boldsymbol{q},\omega_n)|^2  \\ 
-\,\frac{1}{2V^3\beta^3\hbar^4}\sum\limits_{1\,2\,3}
b(1,2,3)\; z(1) z^{\star}(2) z(3) z^{\star}(1-2+3)\bigg) \;.
\end{multline}
Here $Z_0\/$ is the grand partition function of noninteracting 
Fermions while
\begin{equation}
z(\boldsymbol{q},\omega_n)=
\int\limits_{0}^{\beta\hbar}d\tau\; e^{i\omega_n\tau}z(\boldsymbol{q},\tau)
\end{equation}
is the Fourier transform of the $\tau\/$-dependence of 
$z(\boldsymbol{q},\tau)\/$ with bosonic Matsubara frequencies
$\omega_n = 2\pi n/(\beta\hbar)\/$, $n\/$ integer. In the quartic term
we have used the short hand notation $1 = (\boldsymbol{q_1},\omega_1)\/$,
etc. The functions $\tilde{a}\/$ and $b\/$ can be expressed in terms 
of the normal state Green function ($\xi_k^{} = \epsilon_k^{}-\mu\/$, 
$\tilde{\omega}_n =  2\pi (2n+1)/(\beta\hbar)\/$)
\begin{equation}
G_0(\boldsymbol{k},\tilde{\omega}_n)=\frac{1}{i\tilde{\omega}_n-
\xi_{k}^{}/\hbar}
\end{equation}
via
\begin{multline}
\tilde{a}(\boldsymbol{q},\omega_n) = \frac{1}{V\beta\hbar^2}
\sum\limits_{\boldsymbol{k}\,\tilde{\omega}_n}
v_{\boldsymbol{k}-\boldsymbol{q}/2}^{\;2}\,\\ \cdot\,
G_0(\boldsymbol{k}+\boldsymbol{q},\tilde{\omega}_n+\omega_n)
G_0(-\boldsymbol{k},-\tilde{\omega}_n)
\end{multline}
and a similar expression with four factors $G_0\/$ for $b\/$. 
In order to obtain the standard form of a quantum GL functional, the
coefficients $\tilde{a}(\boldsymbol{q},\omega_n)\/$ and 
$b(1,2,3)\/$ have to be expanded for small $\boldsymbol{q}\/$ and
$\omega_n\/$. To lowest order in the spatial and temporal gradients
of the order parameter, it is sufficient to keep only
the leading terms in
\begin{equation} \label{ref17} \hspace*{-0.0in}
a(\boldsymbol{q},\omega_n) = \frac{1}{g} -
\tilde{a}(\boldsymbol{q},\omega_n) = 
a + c\,\frac{\hbar^2 q^2}{2m} - i \,d\,\hbar\omega_n + \cdots
\end{equation}
and replace
\begin{equation}
b(1,2,3) \rightarrow b(0,0,0) = b
\end{equation}
by its constant value at zero momentum and frequency.
This expansion is valid, provided
the contributions of order $\boldsymbol{q}^4\/$ and $\omega_n^{\,2}\/$
in Eq. (\ref{ref17}) are negligible. From an explicit calculation of these
higher order terms it may be shown \cite{stin96} that the order parameter
must vary slowly on length scales of order 
$\xi_b \approx\hbar/\sqrt{mE_b}\/$ with $E_b\/$ the two particle 
binding energy introduced in Eq. (\ref{ref23}) below. 
Physically, the length $\xi_b\/$ is just the radius of a bound state 
with energy $E_b\/$.
In the weak coupling limit this length coincides with
the standard BCS coherence length $\xi_0\approx\hbar v_F/T_c\/$
which, for the clean limit considered here, is identical with the
GL coherence length $\xi(0)\/$ as defined in (\ref{ref49}). 
The standard form of the 
GL functional with a gradient term $|\nabla\psi|^2\/$ is therefore valid
provided the order parameter varies on scales larger than $\xi_b \approx
\xi(0)\/$. With increasing strength of the coupling $E_b\/$, the pair radius
$\xi_b\/$ decreases and thus the validity of the expansion (17)
extends to variations on shorter length scales. Regarding the dependence on
$\tau\/$, the requirement is that $\psi(\boldsymbol{r},\tau)\/$ must
vary slowly on time scales $\tau_b\approx\hbar/E_b\/$. For weak coupling
this is a rather large scale of order $\hbar\epsilon_F/T_c^2\/$ (we set
$k_B=1\/$). Similar to the spatial dependence, however, the necessary scale 
for the $\tau\/$-dependence of the order parameter for which the leading terms
kept in (17) are sufficient, decreases with increasing coupling. Thus, in the
Bose limit, the description of the time dependence of the order parameter by
a first order derivative like in the well known Gross-Pitaevskii
equation \cite{Gros63} becomes exact (see also section VI).

With these approximations our GL functional in
\begin{equation}
Z = Z_0\;{\displaystyle \int} D^2 z\;\exp
(-\beta F[z])
\end{equation}
finally takes the form
\begin{multline} \label{ref20}
F[z] = \frac{1}{\beta\hbar}\int\limits_{V}d^2 r 
\int\limits_{0}^{\beta\hbar}d\tau\bigg(a\,|z(\boldsymbol{r},\tau)|^2 + c\, 
\frac{\hbar^2}{2m}\,|\boldsymbol{\nabla}z(\boldsymbol{r},\tau)|^2 \\
+ \, d\,\hbar\, 
z^{\star}(\boldsymbol{r},\tau)\,\partial_{\tau}z(\boldsymbol{r},\tau) 
+ \frac{b}{2}\, |z(\boldsymbol{r},\tau)|^4\bigg)
\end{multline}
which reduces to the familiar expression if $z\/$ is independent of $\tau\/$.
The coefficients $a\/$ and $b\/$ are given by
\begin{equation} \label{ref21}
a =  \frac{1}{g} - \frac{1}{2V}\sum\limits_{\boldsymbol{k}}
v_{\boldsymbol{k}}^{\,2}\;\frac{\tanh(\beta\xi_k^{}/2)}{\xi_k^{}}
\end{equation}
and
\begin{equation}
b = \frac{1}{V\beta}\sum\limits_{\boldsymbol{k}}\sum\limits_{\omega_n}
\frac{v_{\boldsymbol{k}}^{\,4}}{(\xi_{k}^{\;2}+
\hbar^2\omega_n^{\,2})^2}\;.
\end{equation}
Now the sum over wavevectors $\boldsymbol{k}\/$ in Eq. (\ref{ref21}) diverges
at large $k\/$ and thus the bare value of $a\/$ is undefined.
In the weak coupling limit this divergence is usually
eliminated by argueing that the interaction is finite only in a thin
shell around the Fermi surface. In the present case however, where the 
condensation in the strong coupling limit really affects the whole 
Fermi sphere, such a procedure is no longer possible. Instead, as was 
pointed out by Randeria et al.\cite{RaDuSh90}, we have to connect the bare 
coupling constant to the low energy limit of the two-body scattering
problem. In two dimensions this relation is of the form 
\begin{equation} \label{ref23}
\frac{1}{g} = 
\frac{m}{4\pi\hbar^2}\ln\bigg(\frac{2\epsilon_{\Lambda}}{E_b}\bigg),
\end{equation}
where $\epsilon_{\Lambda} \rightarrow \infty\/$ is a high energy cutoff
which precisely cancels the large $k\/$ divergence on the right hand side of 
Eq. (\ref{ref21}). The parameter $E_b > 0\/$ is the binding energy of the two
particle bound state in vacuum, which in fact must be finite 
in order to obtain a superconducting instability in two dimensions. 
In our present model, which neglects the dependence of 
$V_{\boldsymbol{k}-\boldsymbol{k'}}^{}\/$ on the absolute values
of $\boldsymbol{k}\/$ and $\boldsymbol{k'}\/$,
the existence of a bound state is indeed a necessary
condition for superconductivity even in the case of d-wave pairing,
although quite generally it only applies in the s-wave case \cite{RaDuSh90}.
Since in the effective mass approximation $\epsilon_k=\hbar^2k^2/2m\/$
which we are using throughout, the free Fermion density of 
states in two dimensions is constant, the coefficient $a\/$ 
can now be calculated analytically in terms of $E_b\/$ as
\begin{multline}
a =  -\frac{m}{4\pi\hbar^2}\Bigg( 2\ln (\frac{4e^{\gamma}}{\pi})
\,\theta (\mu) + \ln (\frac{\beta E_b}{4}) +
\ln (\frac{\beta |\mu|}{2})\\ \cdot\,\tanh (\frac{\beta\mu}{2}) 
+\frac{\beta\, {\rm sgn}(\mu)}{2}
\int\limits_{|\mu|}^{\infty}d\xi\;\frac{\ln (\beta\xi /2)}{\cosh^2
(\beta\xi /2)}\Bigg)
\end{multline}
($\gamma=0.577\ldots\/$ is the Euler constant.).
The coefficient $b\/$ in (22) is finite without a cutoff and given by
\begin{multline} \label{ref25}
b = \frac{3m}{16\pi\hbar^2}\Bigg(
\frac{7\zeta (3)}{2\pi^2}\,\beta_c^{\,2}\,\theta (\mu_c) -
\frac{1}{\mu_c^{\,2}}\,\tanh\big(\frac{\beta_c \mu_c}{2}\big) 
\\ +\,{\rm sgn}(\mu_c)
\int\limits_{|\mu_c|}^{\infty}d\xi\;\frac{\tanh (\beta_c \xi/2)}{\xi^3}
\Bigg)\;.
\end{multline}
Here $\theta (x)\/$ and ${\rm sgn}(x)\/$ are the well known Heaviside
and sign functions. Moreover we
we have replaced $\beta\/$ and $\mu\/$ by their values at the 
critical point. Similarly, the values of 
the two remaining coefficients $c\/$ and $d\/$ at the critical point are
\begin{multline} \label{ref26}
c = \frac{m}{8\pi\hbar^2}\Bigg(
\frac{7\zeta(3)}{2\pi^2}\,\beta_c^{\,2}\mu_c\,\theta(\mu_c) \\ +\,
|\mu_c|\int\limits_{|\mu_c|}^{\infty}d\xi\;\frac{\tanh(\beta_c\xi/2)}{\xi^3}
\Bigg)
\end{multline}
and
\begin{equation} \label{ref27}
d = \frac{m}{8\pi\hbar^2}\,\frac{\tanh(\beta_c\mu_c/2)}{\mu_c}\;.
\end{equation}
All four GL coefficients can thus be expressed essentially in analytical
form for arbitrary strength of the interaction.
Comparing these results with those for the s-wave case \cite{DreZwe92}, 
it turns out that up to a geometrical factor $3/2\/$ in $b\/$ the
coefficients are identical at given values of $\beta_c^{}\/$ and
$\mu_c^{}\/$, provided the two-particle binding energy $E_b\/$ is simply 
identified with the corresponding s-wave value.

\section{Order parameter and Nelson-Kosterlitz jump}
In order to relate the formal auxiliary field $z(\boldsymbol{r},\tau)\/$
in the functional (\ref{ref20}) to the usual superconducting order parameter 
$\psi(\boldsymbol{r},\tau)\/$, the standard procedure in weak coupling is to 
take $\psi_{\rm BCS}^{} = \sqrt{2c}\,z\/$, which gives the conventional 
coefficient $\hbar^2/4m\/$ in front of $|\nabla\psi_{\rm BCS}^{}|^2\/$. 
The gradient term is thus
identical with the kinetic energy of a Schr\"odinger field for a single
quantum mechanical particle with mass $m_{\star}=2m\/$, describing a pair
built from constituents with mass $m\/$. 
As pointed out by 
de Gennes \cite{deGe89}, however, the value of $m_{\star}\/$ is arbitrary 
in principle, as long as one is considering the classical
GL functional with $\psi\/$ independent of $\tau\/$. Indeed all measurable
quantities obtained from the classical GL functional depend only on
ratios like $|\psi|^2/m_{\star}\/$. An arbitrary choice for $m_{\star}\/$
can therefore always be compensated by an appropriate rescaling of $\psi\/$.
This situation is changed, however, in a quantum mechanical treatment,
where the order parameter also depends on $\tau\/$, i.e. dynamics enters.
In this case there is a different natural normalization of 
$\psi = \sqrt{d}\, z\/$ in which the coefficient of the
$\psi^{\star}\,\partial_{\tau}\psi\/$-contribution is just $\hbar\/$
\cite{DreZwe92,Zwer92}. With
this choice of normalization, the order parameter
$\psi(\boldsymbol{r},\tau)\/$ is precisely
the c-number field in a coherent state path integral \cite{Schu81} associated
with a genuine Bose field operator $\hat{\psi}(\boldsymbol{r})\/$ 
with canonical commutation relations 
$[\hat{\psi}(\boldsymbol{r}),\hat{\psi}^{\dagger}(\boldsymbol{r'})] 
= \delta(\boldsymbol{r}-\boldsymbol{r'})\/$. 
While this normalization is evidently the most appropriate one in the
strong coupling Bose limit - where it agrees with the standard choice as 
we will see - it can be used for arbitrary coupling, even in the
BCS-limit. Including the charge $-2e\/$ of a pair by generalizing the
gradient to a covariant derivative in the standard way and adding the
energy associated with the magnetic field $\boldsymbol{h}=\boldsymbol{\nabla}
\times\boldsymbol{A}\/$, the resulting free energy functional reads
($c_0\/$ is the velocity of light)
\begin{multline} \label{ref28}
F[\psi,\boldsymbol{A}] = \\
\frac{1}{\beta\hbar}\int\limits_{V}d^2r 
\int\limits_{0}^{\beta\hbar}d\tau\Big(-\mu_{\star}\,
|\psi|^2 + 
\frac{1}{2m_{\star}}\,\Big|\Big(\frac{\hbar}{i}\boldsymbol{\nabla}
+\frac{2e}{c_0}\boldsymbol{A}\Big)\psi\Big|^2 \\
+ \, \hbar\, \psi^{\star}\,\partial_{\tau}\psi
+ \frac{g_{\star}}{2}\, |\psi|^4\Big)  +
\frac{\boldsymbol{h}^2}{8\pi}\;.
\end{multline}
With this normalization, the three remaining indepen\-dent coefficients now
have a very direct physical inter\-pretation \cite{Zwer92}: the coefficient
$\mu_{\star} = -a/d\/$ is the effective chemical potential of the Bosons,
$m_{\star} = m\;d/c\/$ their effective mass and $g_{\star} = b/d^2\/$ a
measure of the repulsive interaction between the composite Bosons. 

\begin{figure}
\epsfig{file=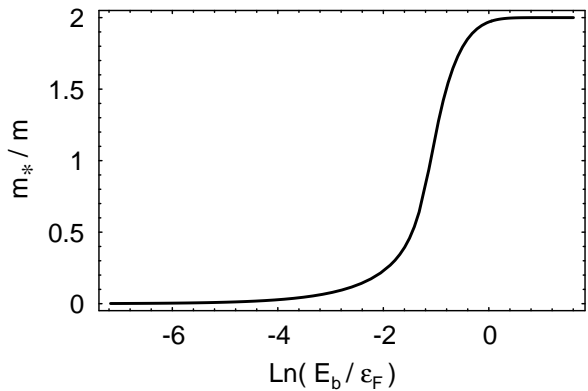,scale=0.55}
\caption{\label{fig1}The effective mass  $m_{\star}/m\/$ of the 
composite Bo\-sons versus the binding energy $E_b/\epsilon_F\/$.}
\end{figure}

In the following we will concentrate on the effective mass $m_{\star}\/$
(at $T_c\/$) which, according to (\ref{ref26},\ref{ref27}) 
is completely determined by
the ratio $T_c/\mu_c\/$. Now in the weak coupling limit $\mu_c\/$ is equal
to the Fermi energy (see section IV below) and thus $m_{\star}/m\/$
vanishes like $(T_c/\epsilon_F)^2\/$ in the case of a BCS-superconductor. 
By contrast, for strong coupling $\mu_c\/$ approaches $-E_b/2\/$, i.e. is
large and negative. In the Bose limit $m_{\star}\/$ is therefore equal 
to $2m\/$ as expected. Relating $T_c/\mu_c\/$ to the dimensionless
coupling strength $\ln{E_b/\epsilon_F}\/$ by the Gaussian approximation
discussed in the following section, we obtain a monotonic increase of
$m_{\star}\/$ from exponentially small values to $2m\/$ as a function of
coupling, as is shown in Fig. \ref{fig1}. 
As was pointed out above, the mass
$m_{\star}\/$ of a Cooper pair defined in such a way cannot be
observed in any static measurement like the penetration depth.
To discuss this, we consider the two dimensional current density 
\begin{equation} \label{ref29} \hspace*{-0.2cm}
\boldsymbol{j}(\boldsymbol{r}) = 
-c_0 \frac{\delta F[\psi,\boldsymbol{A}]}
{\delta\boldsymbol{A}(\boldsymbol{r})}=
-2e|\psi|^2\frac{\hbar}{m_{\star}}\Big(\boldsymbol{\nabla}\phi +
\frac{2e}{\hbar c_0}\boldsymbol{A}\Big)
\end{equation}
which follows from (\ref{ref28}) for a $\tau\/$-independent order parameter
$\psi=|\psi|e^{i\phi}\/$. For a spatially constant magnitude $\psi_{\infty}\/$
of the order parameter, this leads immediately to the London equation
\begin{equation} \label{ref30}
\boldsymbol{\nabla}\times\boldsymbol{j}(\boldsymbol{r}) =
-\frac{4e^2|\psi_{\infty}|^2}{m_{\star}c_0}\boldsymbol{h}\;.
\end{equation}
As was noted above, it is only the ratio $|\psi|^2/m_{\star}\/$ which enters
here and thus static magnetic properties are independent of the choice
for $m_{\star}\/$. Specifically we consider a thin 
superconducting film with thickness $\delta\/$. The in-plane 
penetration depth $\lambda\/$ is then related to $|\psi_{\infty}|^2\/$
via \cite{HN79}
\begin{equation} \label{ref31}
L_s = \frac{2\,\lambda^2}{\delta}=\frac{m_{\star}c_0^{\,2}}
{8\pi|\psi_{\infty}|^2e^2}\;.
\end{equation}
Here we have introduced a further length $L_s\/$ which is the effective
magnetic penetration depth in a thin film. Typically this length is of the
order of one centimeter and thus for sample sizes which are smaller than that,
magnetic screening may be neglected. In such a situation the difference
between a charged and a neutral superfluid becomes irrelevant. A 
superconducting film thus exhibits a Kosterlitz-Thouless transition
\cite{Minn87}, in which the renormalized helicity modulus
$\gamma(T)=\hbar^2|\psi|^2(T)/m_{\star}\/$ jumps from $2T_c/\pi\/$ to
zero at $T_c\/$. Using (31) this jump translates into one for the 
two dimensional screening length $L_s\/$ of size \cite{HN79}
\begin{equation} \label{ref32}
\left.(L_s(T)\;T)\right|_{T_c^{\,-}} = 
\Big(\frac{\phi_0}{4\pi}\Big)^2=1.96\; \text{K\,cm}
\end{equation}
where $\phi_0=hc_0/2e\/$ is the standard flux quantum. Consistent with our
remarks above, this jump is completely universal and independent of 
$m_{\star}\/$, applying both to BCS- or Bose-like superconductors,
provided $L_s\/$ is larger than the sample size and the density of
vortices is low \cite{Minn87}. 

In order to define a proper superfluid density $n_s\/$, we consider
the relation between the order parameter $\psi\/$ and the microscopic
anomalous average. From Eq. (\ref{expect}) we have 
$\psi_{\rm BCS} = -g\sqrt{2c}\, \langle b \rangle\/$. Neglecting the internal
d-wave structure of the order parameter and the logarithmic factor in 
(\ref{ref23}), it is straightforward to see that 
\begin{equation} \label{ordpar}
\psi_{\rm BCS}(\boldsymbol{r}) \approx \xi_b 
\langle\hat\psi_{+1}(\boldsymbol{r})\;
\hat\psi_{-1}(\boldsymbol{r})\rangle
\end{equation}  
with $\hat\psi_{\sigma}(\boldsymbol{r})\/$, $\sigma = \pm 1\/$ the 
Fermionic field
operators (The factor $\ln (2\epsilon_{\Lambda}/E_b)\/$ which is 
omitted in (\ref{ordpar}) diverges as $\epsilon_{\Lambda} 
\rightarrow \infty\/$. This is a reflection of the fact that the 
product of two field operators at the same point can properly defined 
only with a cutoff.). Since $\xi_b\/$ is the radius of a pair, the 
relation (\ref{ordpar}) shows that 
$|\psi_{\rm BCS}(\boldsymbol{r})|^2\/$ is just the areal density of 
pairs. This remains true even in the Bose limit where 
$\xi_b\rightarrow 0\/$ while the product 
$\hat\psi_{+1}(\boldsymbol{r})\,\hat\psi_{-1}(\boldsymbol{r})\/$
eventually behaves like a true Bose field operator 
$\hat\psi(\boldsymbol{r})\/$. The standard definition 
$n_s = 2 |\psi_{\rm BCS}|^2\/$ of the superfluid density can thus 
be applied for arbitrary coupling. By contrast, the Bose order
parameter $\psi = \sqrt{m_{\star}/2m}\,\psi_{\rm BCS}\/$ coincides 
with $\psi_{\rm BCS}\/$ only in the strong coupling limit. For weak 
coupling it is given by an expression like (\ref{ordpar})
but with the interparticle spacing $k_F^{\,-1}\/$ instead of $\xi_b\/$ 
as the prefactor. Thus $|\psi(\boldsymbol{r})|^2\/$ is essentially
the probability density for two Fermions with opposite spin at the 
same point. In the BCS limit this density is exponentially small due 
to the large size of a pair. The superfluid density in turn is still 
of order one even in weak coupling and indeed at zero temperature
$n_s\/$ must be equal to the full density $n\/$ for any superfluid 
ground state in a translational system as considered here \cite{Tony73}.
Using $n_s = 2 |\psi_{\rm BCS}(\boldsymbol{r})|^2 \/$, the relation 
(\ref{ref32}) can be rewritten in terms of a jump
\begin{equation} \label{ref34}
n_s(T_c^-) =
\frac{2m}{\pi\hbar^2}\;T_c
\end{equation}  
of the renormalized superfluid density. The superfluid fraction $n_s/n\/$
therefore has a jump of order $T_c/T_c^{\,\rm Bose}\/$. Since this ratio
approaches zero in weak coupling, there is a smooth crossover between the
universal jump of the superfluid density in a Bose superfluid \cite{NK77}
and the behaviour in a strict BCS model where $n_s/n = 2(1-T/T_c)\/$
vanishes {\em continuously} near $T_c\/$ even in two dimensions. Indeed
the BCS-Hamiltonian is equivalent to a model with an infinite
range interaction of strength $V^{-1}\/$ for which mean field theory
is exact \cite{MS63}. In the following we will neglect the 
subtleties associated with the Kosterlitz-Thouless nature of the
transition, which is anyway masked by the coupling between different
CuO$_2\/$-planes in real high-$T_c\/$ superconductors, giving a
three dimensional critical behaviour near $T_c\/$ \cite{Schn94}. 

\section{Gaussian Approximation}
In order to calculate directly observable quantities from our
GL functional (\ref{ref28}), we have to determine both the critical 
temperature
$T_c\/$ and the corresponding chemi\-cal potential $\mu_c\/$ in terms 
of the binding energy $E_b\/$. 
Now it is obvious that an exact evaluation of the functional integral
over $\psi(\boldsymbol{r},\tau)\/$ is impossible. We will therefore use 
the Gaussian approximation above $T_c\/$, which is obtained by simply omitting
the $|\psi|^4\/$-term. With this approximation our complete grand
canonical potential $\Omega\/$ per volume takes the form
\begin{multline} \label{ref34b}
\Omega = \Omega_0 + \frac{1}{\beta V} 
\sum\limits_{\boldsymbol{q}\,\omega_n}\;\ln\;gd\Big(-\mu_{\star}+ 
\frac{\hbar^2 q^2}{2m_{\star}} - i\,\hbar\omega_n\Big)\,. 
\end{multline}
The critical temperature and chemical potential then follow from
the standard condition
\begin{equation} \label{ref35}
\mu_{\star}(T_c,\mu_c) = 0
\end{equation}
for a bifurcation to a nonzero order parameter, and the particle number
equation
\begin{multline} \label{ref36}
n = - \frac{\partial \Omega}{\partial\mu} = n_0 \\ + \,
\partial_{\mu}\mu_{\star} 
\int \frac{d^2q}{(2\pi)^2}\;\frac{1}{\exp
\big[\beta\big(-\mu_{\star}+\frac{\hbar^2 q^2}{2m_{\star}}\big)\big]-1}\;.
\end{multline}
Here
\begin{equation}
n_0 =
\frac{m}{\pi\hbar^2}\,\frac{\ln(1+\exp(\beta\mu))}{\beta}
\end{equation}
is the number density of a free Fermion gas in two dimensions.
Eq. (\ref{ref35}) is identical with the Thouless criterion 
\cite{Thoule60} for a superconducting instability, which is 
equivalent to the condition that the ladder approximation to the exact 
pair field susceptibility
\begin{equation}
\chi_{pair}=\int\limits_{0}^{\beta\hbar}d\tau\;\langle 
b_{\boldsymbol{0}}^{}(\tau)\,b_{\boldsymbol{0}}^{}(0)\rangle
\end{equation}
diverges \cite{ToMiYa93}. It is a straightforward generalization of
the usual gap equation to arbitrary coupling. The number equation
(\ref{ref36}) deserves some more comments. Since we have
\begin{equation} \label{ref39}
d = - \frac{1}{2}\,\partial_{\mu}a|_{\beta=\beta_c,\mu=\mu_c}
\end{equation}
quite generally, it is easy to see that 
$\partial_{\mu}\mu_{\star} = 2\/$ at $\beta=\beta_c\/$ and 
$\mu=\mu_c\/$. Therefore Eq. (\ref{ref36}) has the simple intuitive 
interpretation that the total number of particles is split into the 
number of free Fermions still present at $(\beta_c,\mu_c)\/$ plus the number
of Fermions already bound together in pairs, whose mean occupation
number is just the Bose distribution. Now formally this distribution
function arises from the summation over the Matsubara frequencies 
$\omega_n\/$ in Eq. (\ref{ref34b}) precisely because our coefficient 
$a(\boldsymbol{q},\omega_n)\/$ has been expanded only to linear order
in $\omega_n\/$. The omission of the higher order terms in this
expansion is therefore connected with neglecting scattering state
contributions, which would give an additional term in Eq. (\ref{ref36}) 
beyond the completely free and fully bound number of Fermions. Such a
contribution is important in the three-dimensional case where 
true bound states exist
only beyond a critical strength of the coupling \cite{NozSch84,SaRaEn93}. 
For our present discussion
of the problem in two dimensions, however, there are only free Fermions or
true bound states. Therefore there is no contribution in Eq. (\ref{ref36})
from scattering states and one expects that the expansion of
$a(\boldsymbol{q},\omega_n)\/$ to linear order in $\omega_n\/$ is reliable
at arbitrary strength of interaction. 

There is however a rather different
problem which appears in the two-dimensional case. As was discussed above,
a superconducting transition exists only in
the Kosterlitz-Thouless sense. This  problem shows up
in our Gaussian approximation, since the Bose integral in Eq. (\ref{ref36})
diverges at $\mu_{\star} = 0\/$. Now at this level of approximation
this is just a reflection of the fact that $T_c=0\/$ for an ideal Bose gas
in two dimensions,
because - as pointed out above - the omission of the $|\psi|^4\/$-term
corresponds to neglecting the repulsive interaction between the Bosons.
From our analytical results (\ref{ref25}) and (\ref{ref27}) for $b\/$ and
$d\/$, which do not contain $E_b\/$, it is however straightforward to 
calculate the effective interaction $g_{\star}\/$ for arbitrary coupling.
It turns out that $g_{\star}\/$ is a monotonically decreasing function
of the coupling.
In the limits $\beta_c\mu_c \rightarrow \infty\/$ of a BCS-like 
or  $\beta_c\mu_c \rightarrow -\infty\/$ of a Bose-like system,
we find
\begin{equation} \label{ref40}
g_{\star} = \frac{6\pi\hbar^2}{m} \begin{cases}
\frac{\textstyle 7\zeta (3)}{\textstyle \pi^2}\, (\beta_c\mu_c)^2
& \text{BCS} \\
1 & \text{Bose}
\end{cases}\,\,.
\end{equation}
Thus, in two dimensions, there is always a finite repulsive 
interaction between the pairs,
which is of purely statistical origin \cite{DreZwe92,Zwer92}.
In particular $g_{\star}\/$ remains finite in the Bose limit, where it 
arises from processes with a virtual exchange of one of the constituent 
Fermions in a Bose-Bose scattering process \cite{Haussm93}.
The fact that $g_{\star}\/$ is very large in the weak coupling limit is
simply a consequence of the large pair size 
$\xi_0 \sim \beta_c\epsilon_F\/$ (note that 
$\mu_c = \epsilon_F\/$ in the weak coupling limit), but does not imply
that the $|\psi|^4\/$-term is particularly relevant in this regime.
On the contrary, using the Gaussian approximation, 
it is straightforward to show \cite{stin96} that in this
limit the pro\-duct $g_{\star}\langle |\psi|^2 \rangle\/$ which
effectively renormalizes the Boson chemical potential $\mu_{\star}\/$, is
of order $\sqrt{E_b\epsilon_F}\/$ which is roughly $T_c\/$ in the weak
coupling limit. For BCS-like superconductors the 
$|\psi|^4\/$-contribution is therefore irrelevant except very close to 
$T_c\/$, a fact which is well known from the
standard theory of conventional superconductors. Now the finite
value of $g_{\star}\/$ guarantees that even in two dimensions
there is a finite critical temperature below which the superfluid density
is nonvanishing. 
Unfortunately it is not possible to incorporate the  
Kosterlitz-Thouless nature of the transition in an approximate
treatment of the GL functional. However considering the effectively
three-dimensional structure of high-$T_c\/$ superconductors, this problem 
may be circumvented by including the motion
in the direction perpendicular to the planes. A very simple method to
incorporate this is provided by adding a transverse contribution
$\epsilon_{\perp}\/$ to the kinetic energy by \cite{GuLoSh95}
\begin{equation} \hspace*{-0.5in}
\frac{\hbar^2 q^2}{2m_{\star}} \rightarrow \frac{\hbar^2 q^2}{2m_{\star}} +
\frac{\hbar^2 q_{\perp}^{\,2}}{2m_{\perp}}\;,
\end{equation} 
whose average is equal to the thermal energy $\langle\epsilon_{\perp}
\rangle =T\/$.
Replacing the integral in Eq. (\ref{ref36}) by
\begin{equation}
\int\frac{d^2q}{(2\pi)^2}\;\;\rightarrow L_{\perp}\;\,
\int\frac{d^2q}{(2\pi)^2} \int\frac{dq_{\perp}}{2\pi}
\end{equation} 
with $L_{\perp} = 2\pi/ \sqrt{\langle q_{\perp}^{\,2}\rangle}\;\/$
we obtain an effectively 
three-dimensional system. This becomes evident by writing 
the contribution of the bound pairs in (\ref{ref36}) in the form
\begin{equation} \label{ref43} \hspace*{-1.0cm}
n' = \frac{m_{\star}\,\partial_{\mu}\mu_{\star}\,\sqrt{\beta}}{\pi\hbar^2}\,
\int\limits_0^{\infty}d\epsilon
\;\frac{\sqrt{\epsilon}}{\exp[\beta(-\mu_{\star}+\epsilon)]-1}\;.
\end{equation}
The density of states is thus proportional to $\sqrt\epsilon\/$ as in
three dimensions, making $n'\/$
finite at $\mu_{\star} = 0\/$. Although rather crude, this
approximation gives a value for $T_c\/$ in the Bose limit, which is
very close to the Kosterlitz-Thouless value for the transition 
temperature \cite{Minn87}
\begin{equation} \label{ref44}
T_c^{KT} = {0.89} \frac{\pi\hbar^2n}{4m k_B}
\end{equation}
of a dilute hard core Bose gas on a lattice\cite{FisHoh87}
with Boson mass $2m\/$ and number density $n/2\/$.

Using our results for the GL coefficients $a\/$, $c\/$ and $d\/$
and the replacement (\ref{ref43}) in the number equation, we can now 
determine both $T_c\/$ and $\mu_c\/$ for arbitrary coupling from 
$\mu_{\star}=0\/$ and Eq. (\ref{ref36}). The corresponding results
are shown in Figs. 2 and 3 
in units of the characteristic energy scale $\epsilon_F\/$ and
as a function of the dimensionless 
effective coupling $E_b/\epsilon_F\/$.
For weak coupling $E_b \ll \epsilon_F\/$, the critical temperature is
monotonic in $E_b\/$, behaving like $T_c \approx
\sqrt{E_b\epsilon_F}\/$. For intermediate coupling, it exhibits a 
maximum \cite{DreZwe92}. A similar but less pronounced 
behaviour is found in three dimensions, again in the  
Gaussian approximation \cite{NozSch84,SaRaEn93}.
In a more refined 
self-consistent treatment \cite{Haussm94},
however, the critical temperature is a monotonically increasing function
of the coupling.
It is likely that the same situation 
also applies in two dimensions, but unfortunately there is at present no
quantitative theory taking into account the repulsive interaction between
the pairs in this case. In the Bose limit
the transition temperature becomes independent of the original attractive
interaction and is completely determined by the  Boson density 
$n/2\/$ and the effective mass $m_{\star} \rightarrow 2m\/$.  
It approaches a value of about $10\%\/$ of the Fermi
energy, which is likely to be an upper limit for $T_c\/$ in the present
problem. 
\begin{figure}
\epsfig{file=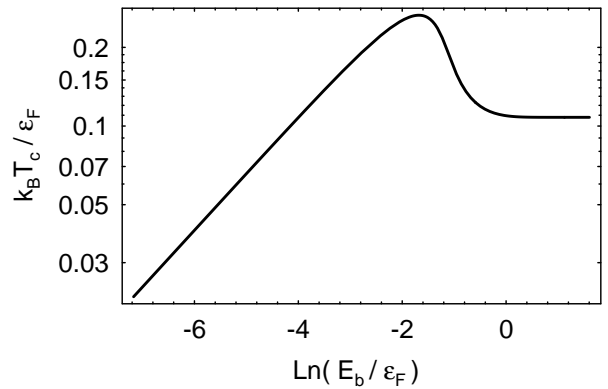,scale=0.55}
\caption{\label{fig2}The normalized critical temperature $T_c\/$ versus 
the dimensionless coupling strength.}
\end{figure}
The chemical potential $\mu_c\/$ at the transition decreases
monotonically from its weak coupling value $\epsilon_F\/$ to $-E_b/2\/$
in the Bose limit. It changes sign at 
$\ln(E_b/\epsilon_F) \approx -1.1\/$, where the behaviour crosses over from 
BCS- to Bose-like (In Fig \ref{fig3} we have supressed a small dip
in $\mu_c\/$ around these couplings which is an artefact of the
pronounced maximum in $T_c\/$).
Apart from the chemical potential $\mu_c\/$, 
the nature of the transition can also
be inferred from evaluating the number of preformed Bosons  at $T_c\/$.
This quantity, which is just half of the contribution (\ref{ref43}) 
to the number equation is shown in Fig. \ref{fig4}.
It is obvious that the nature of the phase transition changes rather
quickly in a range of couplings between $\ln(E_b/\epsilon_F) \approx -2\/$
and $-1\/$.  For 
smaller couplings, the density of preformed pairs near $T_c\/$ is
essentially negligible and binding occurs simultaneously with 
condensation. On the other hand, for 
$\ln(E_b/\epsilon_F)\gtrsim -1\/$ basically all pairs are already
present above $T_c\/$ and the transition to superconductivity 
is that of a true Bose system.
\begin{figure}
\epsfig{file=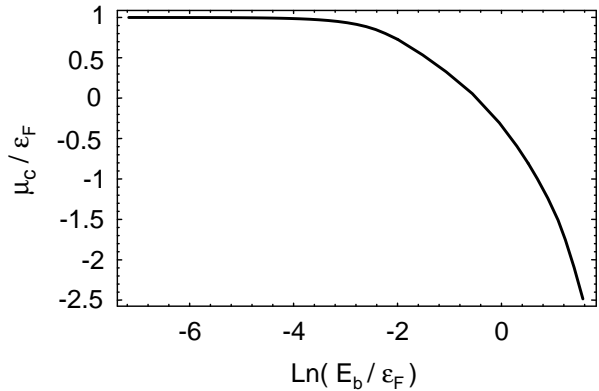,scale=0.55}
\caption{\label{fig3}The normalized chemical potential $\mu_c\/$ at the 
critical point versus coupling.}
\end{figure}
\begin{figure}
\epsfig{file=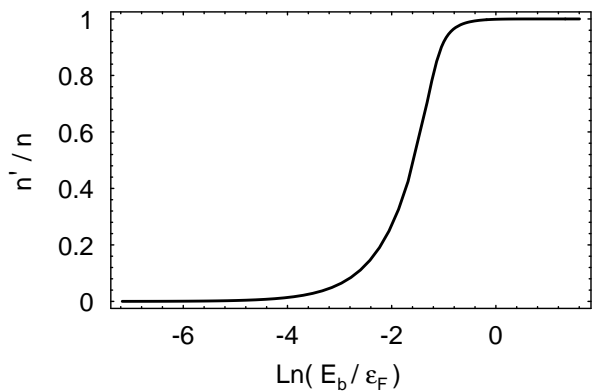,scale=0.55}
\caption{\label{fig4}Contribution $n'\/$ of bound pairs 
at the critical point to the total Fermion number density $n\/$.}
\end{figure}

\section{Characteristic Lengths}
\begin{figure}
\epsfig{file=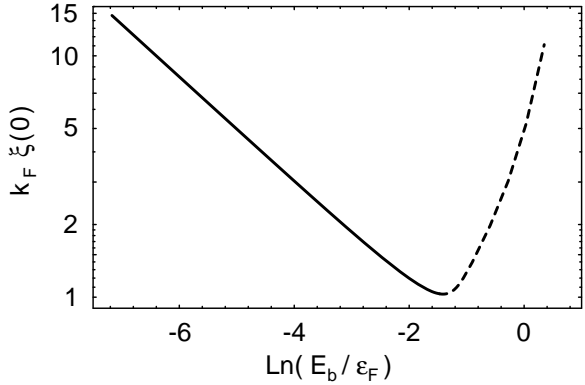,scale=0.55}
\caption{\label{fig5}The Ginzburg-Landau coherence length $\xi(0)\/$ versus
the coupling strength.}
\end{figure}

\begin{figure}
\epsfig{file=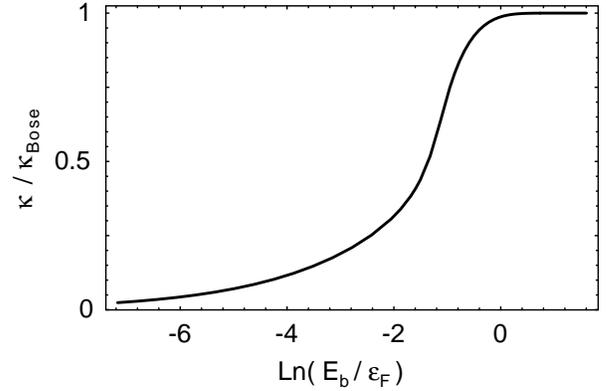,scale=0.55}
\caption{\label{fig6}The GL parameter $\kappa\/$ versus
coupling.}
\end{figure}

In the following we want to determine both the coherence length $\xi\/$
and the penetration depth $\lambda\/$ as a function of the coupling. The
former is defined both above and below $T_c\/$ and may be obtained from
the GL coefficients simply via
\begin{equation} 
\xi^2 = \frac{\hbar^2}{2m}\frac{c}{a}\;.
\end{equation} 
The definition of a penetration depth in a two-di\-mensional 
super\-conductor has been discussed in section III. Within the Gaussian
approximation we may replace $|\psi_{\infty}|^2\/$ in (31) by 
$\mu_{\star}/g_{\star}\/$ which leads to
\begin{equation}
\lambda^2=\lambda_L^2\cdot\frac{bn}{4c|a|}\;.
\end{equation}
Here we have introduced the bare value $\lambda_L\/$ of the London 
penetration depth defined by
\begin{equation}
\lambda_L^2=\frac{mc_0^{\,2}}{4\pi n_3e^2}
\end{equation}
with $n_3=n/\delta\/$ the nominal three-dimensional carrier density. Note
that both $\xi\/$ and $\lambda\/$ have been written in terms of the original
static GL-coefficients $a,b\/$ and $c\/$, in order to stress that the
characteristic lengths are independent of the normalization of $\psi\/$,
i.e. the kinetic coefficient $d\/$ does not enter here. Since the
coefficient $a\/$ vanishes at the transition, both $\xi\/$ and 
$\lambda\/$ diverge. Now in a full treatment of the GL-functional,
including the $\psi^4\/$-term, the behaviour very close to $T_c\/$
in a single layer would be of the Kosterlitz-Thouless type. The
correlation length would thus diverge like \cite{Minn87}
$\ln{\xi}\sim |T-T_c|^{-1/2}\/$ while $\lambda^{-2}\/$ would jump from zero
to a finite value below $T_c\/$ (see (32)). In the 
three-dimensional case, the behaviour very close to $T_c\/$ is that
of a 3d XY-model with nontrivial 
but well known critical exponents \cite{Schn94}.
In our Gaussian approximation this complex structure is replaced by a
simple mean field behaviour. However there is a subtle point even at this
level of approximation. Indeed the coefficient $a\/$ depends both on
temperature and chemical potential and it is only in the strict
BCS-limit, where the latter is a fixed constant equal to $\epsilon_F\/$.
With increasing coupling, however, the chemical potential changes and thus
the relevant limit close to $T_c\/$ is to consider
$a(T,\mu(T))\/$ as $T\to T_c\/$. Now by using the exact relation 
(\ref{ref39}), it is straightforward to show \cite{stin96} that
$a(T,\mu(T)) \sim (T-T_c)^2\/$ vanishes quadratically near $T_c\/$. The
resulting critical exponent for the correlation length would thus be
$\nu=1\/$. Indeed this is the exponent expected for an ideal Bose gas in
three dimensions, to which our Gaussian approximation is effectively
equivalent. Now in order to allow a comparison of our results with
measured values of $\xi\/$ and $\lambda\/$, which are found to obey
a mean field behaviour with $\nu=1/2\/$ 
except very close to $T_c\/$ \cite{kama94},
we neglect the temperature dependence of the chemical
potential near $T_c\/$. As a result
\begin{equation}
a(T,\mu_c)=a'\cdot\frac{T-T_c}{T_c}
\end{equation}
vanishes linearly near $T_c\/$, giving the standard mean field divergence
of $\xi\/$ and $\lambda\/$. Obviously this approximation is only reliable
on the weak coupling side of the transition. As we will see below, however,
this is indeed the relevant regime even in high-$T_c\/$ superconductors.
We thus expect that our results are at least qualitatively reliable for
these systems. For strong coupling, the derivative $a'\/$ of $a(T,\mu_c)\/$
near $T_c\/$ vanishes like $\exp(-\beta_c E_b/2)\/$. 
As a result, the
characteristic lengths $\xi\/$ and $\lambda\/$ would increase exponentially
in the Bose limit which is unphysical however. We have therefore restricted
our calculation of the GL coherence length $\xi(0)\/$ defined by
\begin{equation} \label{ref49}
\xi^2(0)=\frac{\hbar^2}{2m}\frac{c}{a'}
\end{equation}
to coupling strengths $\ln (E_b/\epsilon_F)\/$ smaller than about $-1\/$, 
where the system starts to cross over to Bose like behaviour. The corresponding 
result for $\xi(0)\/$ in units of $k_F^{-1}\/$ is shown in Fig. \ref{fig5}. 
It exhibits the expected decrease of the coherence length from 
its weak coupling limit
\begin{equation}
k_F\xi(0)|_{\rm BCS}^{} =  \sqrt{\frac{7\,\zeta(3)}{8\pi^2}}\;
\frac{\epsilon_F}{T_c}
\end{equation}
to values of order one near the crossover regime, before it starts to rise
again. As was noted above our approximations in determining
$\xi(0)\/$ are no longer reliable in this regime. While in three
dimensions $\xi(0)\/$ is expected to increase like $1/\sqrt{a_B}\;\/$
\cite{Wen90} with $a_B=2a_F\to 0\/$ the relevant Bose-Bose scattering
length \cite{SaRaEn93}, the actual behaviour in two dimensions is
unknown. Fortunately, however, this problem does not arise in determining
the GL-parameter $\kappa=\lambda/\xi\/$ which, in two dimensions,
can be expressed as
\begin{equation} \label{ref52}
\kappa=\left(\frac{m}{8\pi\hbar^2}\cdot
\frac{b\,\delta}{c^2 r_{\rm cl}}\right)^{1/2}\;.
\end{equation}
Here we have introduced the equivalent of the classical electron radius 
$r_{\rm cl}=e^2/mc_0^{\,2}\/$ where $m\/$ is the band mass.
Since the problematic coefficient
$a\/$ has dropped out in $\kappa\/$, we can use (\ref{ref52}) to determine the
GL-parameter in the whole regime between weak and strong coupling. In the 
BCS-limit $\kappa\/$ is exponentially small, behaving like
$\kappa_{\rm BCS}^{}\approx T_c/\epsilon_F
\cdot(\delta/r_{\rm cl})^{1/2}\/$. 
The associated penetration depth $\lambda(0) = \sqrt{3/4}\,\lambda_L\/$
is thus essentially equal to the London value $\lambda_L\/$.
By contrast, for the Bose case $\kappa\/$
approaches a constant value
\begin{equation}
\kappa_{\rm Bose}^{} = \Big( \frac{3\delta}{r_{\rm cl}}\Big)^{1/2}
\end{equation}
which is large compared to one, since $\delta\gg r_{cl}\/$ for realistic
values of the sheet thickness $\delta\/$. The complete dependence
of $\kappa\/$ in the crossover regime
is shown in Fig. \ref{fig6}. 
It is a monotonic function of the binding energy.
Thus with increasing coupling there is always a transition from
type I to type II behaviour in two dimensions, even for a clean 
superconductor with no impurities as considered here.

With these results, we are now in a position to compare our simple model
with experimental values for optimally doped high-$T_c\/$ superconductors.
Since the critical temperature is exponentially sensitive to the 
coupling strength and also is likely to be considerably reduced compared
to our result in Fig. 2 by fluctuation effects in the crossover regime,
we have refrained from 
taking $T_c\/$ as a reliable parameter for adjustment. Instead we have 
used the measured values of the slope of the upper critical field
$H_{c_2}\/$ near $T_c\/$, which determines the GL
coherence length $\xi(0)\/$ via 
\begin{equation}
\left. \frac{dH_{c_2}}{d\ln T}\right|_{T=T_c} = 
\frac{\phi_0}{2\pi\xi^2(0)}\;. 
\end{equation}
In order to fix the dimensionless
coupling strength $E_b/\epsilon_F\/$ from the measured values 
\cite{WeKwCr89} $\xi(0) = 14\, \mbox{\AA}\/$ for optimally doped YBCO 
($T_c = 92\,K\/$) and \cite{MaTaOg92} 
$\xi(0) = 24\, \mbox{\AA}\/$ for the corresponding
compound BSCCO ($T_c = 107\,K\/$), we need the in plane carrier densities
$n\/$ which determine an effective value of $k_F = \sqrt{2\pi n}\/$ (It 
should be pointed out that $k_F\/$ is introduced here only as a measure 
of the carrier density. In fact due to the strong attractive interaction,
the actual momentum distribution of the Fermions just above $T_c\/$ will be
far from the standard Fermi distribution, except in the BCS limit.).
The three-dimensional carrier densities in optimally doped 
YBCO and BSCCO  are \cite{SeIsMa91}  $n_3 = 4\cdot 10^{21}\,cm^{-3}\/$ 
and \cite{HarMil92} $n_3 = 2\cdot 10^{21}\,cm^{-3}\/$ 
respectively.
With the corresponding
values \cite{HarMil92}
$\delta = 5.84\,\mbox{\AA}\/$ and $\delta = 9.27\,\mbox{\AA}\/$
of the effective sheet thickness, the resulting Fermi momenta are
$k_F = 0.38\,\mbox{\AA}^{-1}\/$ for YBCO and 
$k_F = 0.34\,\mbox{\AA}^{-1}\/$ for BSCCO. The dimensionless ratio 
$k_F\xi(0) = 5\/$ and $8\/$ between the coherence length and the average 
interparticle spacing then allows us to determine the effective coupling
strength. From Fig. \ref{fig5} we find that $\ln(E_b/\epsilon_F)\/$ is
equal to $-5.1\/$ for YBCO and $-6.0\/$ for BSCCO. As Figs. \ref{fig3}
and \ref{fig4} show, these coupling strengths describe superconductors
which are still on the weak coupling side of the crossover from BCS to
Bose behaviour. For instance the density of preformed Bosons at $T_c\/$
is less then 1\% in both cases (see Fig. \ref{fig4}).
In order to check whether our description is consistent, we determine
the associated values of the GL parameter $\kappa\/$ near $T_c\/$.
From the above values of $\delta\/$ and
the band masses \cite{SeIsMa91} $m =3.5\,m_e\/$ for YBCO and 
\cite{HarMil92} $m =4.4\,m_e\/$ for BSCCO 
($m_e\/$ is the free-electron mass.)
we find that $\kappa_{\rm Bose}^{}\/$ is equal to $1.5\cdot 10^3\/$
and $2.1\cdot 10^3\/$ respectively for the two compounds considered here.
From Fig. \ref{fig6} we thus obtain $\kappa = 99\/$ and $\kappa = 87\/$
for optimally doped YBCO and BSCCO. These numbers agree very well
with the experimentally determined values of $\kappa\/$, which are
\cite{KrGrHo89} $\kappa_{\rm exp} = 100\/$ and \cite{MaTaOg92,HarMil92}
$\kappa_{\rm exp} = 86\/$. Our 
simple one parameter model therefore gives a consistent quantitative
description of the characteristic lengths $\xi\/$ and $\lambda\/$.

\section{Conclusion}
To summarize, we have studied the crossover in the superconducting 
transition between BCS- and Bose-like behaviour within a GL description.
It has been found that optimally doped high-$T_c\/$ superconductors are still
on the weak coupling side of this crossover, although they are certainly
far away from the BCS-limit. Our microscopic model is characterized by a 
single dimensionless parameter, similar to the familiar BCS-Hamiltonian.
While the GL functional has the standard form for arbitrary coupling its
coefficients depend strongly on the coupling strength. The crossover, 
at least in two dimensions, is essentially identical for the s- or
d-wave case. For static properties, only two of the relevant coefficients
$a\/$, $b\/$ and $c\/$ are independent, since the normalization of the order
parameter is arbitrary. Fixing $\xi(0)\/$ from experiment therefore leaves
only one further parameter -- for instance $\kappa\/$ -- as an independent
predicted quantity. The good agreement of $\kappa\/$ with measured values
supports our conclusion that the optimally high-$T_c\/$ compounds are
intermediate between BCS and Bose behaviour. Since the crossover regime
is rather narrow, however, weak coupling theories are still a reasonable
approximation for the relevant coupling strengths. This is consistent
with the empirical fact that a weak coupling approach apparently works 
well in many cases.

Evidently there are a number of important open questions. They may be 
divided into two classes: the first one concerns the problem of a better 
and more complete treatment of our model itself. The second class is related
to the problem, to which extent this model is applicable to  high-$T_c\/$
compounds and what are the necessary ingredients for a more realistic
description. Regarding our microscopic Hamiltonian as a given model, it is 
obvious that our treatment of the associated GL phenomenology is still
incomplete. In particular the Gaussian approximation is obviously
not sufficient to give a quantitatively reliable result for $T_c\/$ at
intermediate coupling. Moreover, the behaviour of the characteristic 
lengths $\xi\/$ and $\lambda\/$ in the strong coupling limit is completely 
unknown. In order to go beyond the Gaussian approximation, it is necessary
to include the pair interaction (i.~e. the $|\psi|^4\/$-term) properly.
Progress in this direction has been made in the three-dimensional
case by Haussmann \cite{Haussm94} and very recently by Pistolesi
and Strinati \cite{PisStr96}. Using a self-consistent and conserving
approximation for the Green- and vertex functions, 
Haussmann obtained a smooth increase of $T_c\/$ with coupling, thus
eliminating the unphysical maximum in the crossover regime found in the 
Gaussian approximation. This approach is rather different from our present 
one and requires extensive numerical work. Pistolesi and Strinati have
performed an essentially analytical calculation of the correlation
length $\xi_0\/$ at zero temperature, using a loop expansion in a functional
approach  similar to our present one. They have shown that the pair radius
$\xi_b \approx \hbar/\sqrt{mE_b}\/$ coincides with $\xi_0\/$ not only 
in the BCS-limit, but down to values around $k_F\xi_0 \approx 10\/$.
Similar to our results for the GL coherence length $\xi(0)\/$ in Fig.
\ref{fig5}, $k_F\xi_0\/$ reaches a minimum of order one in the crossover
regime before it starts to rise again. Since in three dimensions the
behaviour at strong coupling is that of a weakly interacting Bose gas with
scattering length $a_B = 2a_F \rightarrow 0\/$ \cite{SaRaEn93,Haussm93},
$\xi_0\/$ eventually increases like $k_F^{\,-1}/\sqrt{k_Fa_B}\/$ while
$\xi_b \approx a_B\/$ approaches zero \cite{SaRaEn93}. Unfortunately 
for the two-dimensional case, where the Boson interaction $g_{\star}\/$
is finite even at very strong coupling \cite{DreZwe92,Zwer92}, the 
Kosterlitz-Thouless nature of the transition makes it very difficult
to improve upon the simple approximations used here. A first step in this
direction was taken by Traven\cite{traven}. He showed that interactions
between the pair fluctuations guarantee a nonvanishing superfluid density
at finite temperature, in agreement with our arguments below 
Eq. (\ref{ref40}).
However there seems to be no quantitative calculation of the coherence length
in a two-dimensional Bose-like regime even near zero temperature.

A different problem we want to mention here is that of the proper time
dependent GL theory. Since our quantum GL functional is derived from a
microscopic Hamiltonian, in principle it contains the complete information
about the order parameter dynamics, at least as far as intrinsic effects
are concerned. Neglecting higher order terms in the expansion in
$\omega_n\/$, the resulting equation of motion for the order parameter 
in real time $t\/$ is \cite{SaRaEn93}
\begin{equation} \label{time}
d\cdot i\hbar\frac{\partial \psi(\boldsymbol{r},t)}{\partial t} =
\frac{\delta F[\psi]}{\delta\psi^{\star}(\boldsymbol{r},t)}\;.
\end{equation}
Due to the analytic continuation, the coefficient $d = d_1 + i d_2\/$
has now acquired a finite imaginary part $d_2 > 0\/$ which describes
irreversible relaxation. For a better comparison with the standard literature,
it is convenient to choose the conventional order parameter 
$\psi_{\rm BCS} = \sqrt{2c}\,z\/$, where the prefactor of the 
gradient term is $\hbar^2/4m\/$. With this choice of normalization,
the kinetic coefficient $d_1 = m^{\star}/2m\/$ is identical with the
effective mass discussed in section III. It is then evident from Fig.
\ref{fig1} that a Gross-Pitaevskii-like dynamics where $d_1 =1\/$,
is only valid in the Bose limit, while $d_1 \sim (T_c/\epsilon_F)^2\/$
is exponentially small for weak coupling. Indeed for BCS-like systems it is 
$d_2\/$ which is dominant, being of order $T_c/\epsilon_F\/$ in the 
three-dimensional case \cite{Haussm94}. This result reflects
the fact that for weak coupling superconductors the order parameter
dynamics is purely relaxing. Going beyond the BCS-limit, the
associated kinetic coefficient $d_2\/$ has only been evaluated in 
three dimensions \cite{Haussm94}, where it exhibits a maximum at intermediate
coupling. Its behaviour in the Bose-limit and in two dimensions in general,
however, is completely unknown. Since the incorporation of scattering states
in the three-dimensional case requires to go beyond the linear expansion
in $\omega_n\/$, it is likely however, that a simple first order equation like
Eq. (\ref{time}) is in fact not appropriate for describing the dynamics at 
intermediate coupling. It is only at very low temperatures where the situation 
becomes simple again. Indeed from quite general arguments \cite{Stone95},
the dynamics as $T \rightarrow 0\/$ is expected to be of the
Gross-Pitaevskii form, irrespective of the strength of the coupling. Finally
we mention that a proper microscopic
calculation of the complex coefficient $d\/$ is
relevant for understanding the Hall effect in high-$T_c\/$
compounds \cite{Dors92}.

Concerning the question to which extent our model is really applicable to
high-$T_c\/$ superconductors, it is obvious that most of the complexity of
these systems is neglected here. In particular we have assumed that the 
normal state is characterized by a given density of effective mass Fermions
with some instantaneous attractive
interaction \cite{pietr}. Such a system will
certainly be very different from a conventional Fermi liquid for
intermediate or strong coupling. Our conclusion that we are still on the weak
coupling side of the crossover, with a negligible density of preformed
pairs, is consistent with the phenomenology of optimally (and perhaps
overdoped) high-$T_c\/$ superconductors, however it is certainly
inappropriate for the underdoped cuprates. Indeed these systems exhibit 
a gap far above $T_c\/$ which may be interpreted in terms of preformed Bosons.
A GL description of underdoped compounds was very recently developed
by Geshkenbein, Ioffe and Larkin \cite{GeIoLa97}. Assuming that Bosons
form far above $T_c\/$ only in parts of  the Fermi surface and coexist with
unpaired Fermions through $T_c\/$, they obtain a reasonable description
of the phenomenology of certain underdoped materials. This behaviour is quite
different from that obtained in our model, which completely neglects
any effects of the Coulomb repulsion
and band structure. It is obvious
that for a quantitative description of high-$T_c\/$ superconductors both
Coulomb correlations and band structure effects have to be included, which
requires to use lattice models. They provide a quantitative description of 
these complex systems at least in the normal state and allow the calculation
of microscopic properties like spectral functions, etc \cite{Dopf92}.
Unfortunately with these models it still seems impossible to really explain
how d-wave superconductivity arises from the strongly spin and charge
correlated normal state. As a result, effective models like the
negative-U Hubbard model are often used to discuss microscopic properties
of high-$T_c\/$ compounds \cite{Rand92,Micnas95}.
Our present approach is more phenomenological, starting from a model in which
all microscopic details are neglected except for the fact that we have a 
strong pairing interaction in a system of Fermions with given density.
The advantage of such an approach is that it allows a simple calculation
of the relevant lengths $\xi\/$ and $\lambda\/$ and quantities
following from that like the critical fields. The fact that the
resulting GL theory gives a consistent description of optimally doped
cuprates indicates that at least at this level the microscopic details
are not relevant. Certainly our results supporting this view are quite
limited so far and it is necessary to investigate this further. Since the
coefficients of the GL functional near $T_c\/$ are quite generally
determined by the properties in the {\em normal} state, an interesting
future direction would be to see whether superconducting properties
below $T_c\/$ can quantitatively be obtained from the GL functional by
properly incorporating the anomalous behaviour in the normal state.

\acknowledgments
One of the authors (W. Z.) would like to thank A. J. Leggett for his
hospitality at the University of Illinois where this work was completed
and for useful discussions. Part of this work was supported by a grant 
from the Deutsche Forschungsgemeinschaft (S. S.).


\begin{references}
\bibitem{FisHoh87} In the continuum the result (\ref{ref44}) is modified
 by a double logarithmic factor, which is neglected here however, see
 D. S. Fisher and P. C. Hohenberg,
 Phys. Rev. B {\bf 37}, 4936 (1987).
\end{references}
\end{document}